\documentclass[aps,Phys. Rev. Lett.,a4paper,twocolumn,superscriptaddress,citeautoscript]{revtex4-1}
\usepackage[english]{babel}
\usepackage[T1]{fontenc}
\usepackage{amsmath}
\usepackage{hyperref}
\usepackage{units}
\usepackage[charter]{mathdesign}
\usepackage[pdftex]{graphicx}
\usepackage{color}
\begin{document}
\title{A possible mechanism for superconductivity in doped SrTiO$_3$}
\author{D.~van der Marel}\email{dirk.vandermarel@unige.ch}
\affiliation{Department of Quantum Matter Physics, University of Geneva, 24 Quai Ernest-Ansermet, 1211 Geneva 4, Switzerland}
\author{F.~Barantani}
\affiliation{Department of Quantum Matter Physics, University of Geneva, 24 Quai Ernest-Ansermet, 1211 Geneva 4, Switzerland}
\author{C.~W.~Rischau}
\affiliation{Department of Quantum Matter Physics, University of Geneva, 24 Quai Ernest-Ansermet, 1211 Geneva 4, Switzerland}
\date{\today}
\begin{abstract}
The soft ferro-electric phonon in SrTiO$_3$ observed with optical spectroscopy has an extraordinary strong spectral weight which is much stronger than expected in the limit of a perfectly ionic compound. This "charged phonon" effect in SrTiO$_3$ is caused by the close-to-covalent character of the Ti-O ionic bond and implies a strong coupling between the soft ferro-electric phonon and the inter band transitions across the 3 eV gap of SrTiO$_3$. We demonstrate that this coupling leads, in addition to the charged phonon effect, to a pairing interaction involving the exchange of two transverse optical phonons. This process owes its relevance to the strong electron-phonon coupling and to the fact that the interaction mediated by a single transverse optical phonon vanishes at low electron density. We use the experimental soft phonon spectral weight to calculate the strength of the bi-phonon mediated pairing interaction in the electron doped material and show that it is of the correct magnitude when compared to the experimental value of the superconducting critical temperature.
\end{abstract}
\maketitle
The nature of superconductivity in SrTiO$_3$~\cite{schooley1964,koonce1967,bednorz1988} is exceptional for a number of reasons: superconductivity occurs at extremely low carrier densities down to $10^{17}$ cm$^{-3}$ ~\cite{lin2013,collignon2019}, the material is close to a ferroelectric instability, and can be pushed into the ferroelectric state by appropriate Ca or $^{18}O$ substitution~\cite{mueller1979,itoh1999,rowley2014}, with a coexistence of superconductivity and a ferro-electric type symmetry breaking~\cite{stucky2016,rischau2017,tomioka2019}. While the pairing mechanism is believed to have to do with electron-phonon coupling in some form~\cite{cohen1964,eagles1969,appel1969,zinamon1970,ngai1974,jarlborg2000,marel2011,klimin2012,edge2015,gorkov2016,ruhman2016,kedem2016,kedem2018,woelfle2018,rowley2018,ruhman2019}, the question as to the exact nature of the electron-phonon interaction responsible for the pairing has not received a clear answer yet. As a result of the low carrier density the Fermi temperature is low. Consequently $\epsilon_F$ is of the same range or smaller than the phonon frequencies, placing the coupling to the LO mode at 100 meV - to which the coupling is strongest - in the anti-adiabatic limit~\cite{eagles1969,devreese2010,meevasana2010,stucky2016}. Furthermore, the smallness of the Fermi surface has the consequence of drastically suppressing the available phase space for electron-phonon interaction, and suppressing the conventional pairing mechanism mediated by acoustic phonons. Alternative pairing schemes include exchange of intravalley phonons~\cite{cohen1964}, two-phonon exchange~\cite{ngai1974} and longitudinal optical phonons~\cite{gorkov2016}. However, the LO-phonon mediated electron-electron interaction in the static limit is insufficient to describe superconductivity in strontium titanate, and an alternative approach has been explored based on the full dynamical dielectric function~\cite{klimin2012}. Much attention has been drawn by recent ideas on pairing mediated by the ferroelectric soft mode, and the effects on the pairing amplitude when the system approaches the quantum critical point of the ferroelectric order parameter. The quantum critical fluctuations of the ferroelectric order are often believed to be particularly good candidates for mediating a pairing interaction. However, one problem looms over this approach: the soft ferroelectric phonon (TO1) is a transverse optical mode. Processes whereby a $d$-electron emits a TO phonon have vanishing amplitude in the long wavelength limit~\cite{ruhman2019}, which, due to the small Fermi momentum of these most dilute superconductors, is the relevant range. The corresponding longitudinal mode (LO1) has a frequency of about 20 meV, independent of temperature, and remains at this high energy through the phase transition (see Appendix). Another noteworthy feature is the fact that the electron-phonon coupling as described by the Frohlich model is negligible ($\alpha=0.02$) for the LO1 phonon (see Appendix).  This then raises the question whether there is any relevance at all of the soft ferroelectric modes in relation to the superconducting pairing mechanism. 

\begin{figure}[t!!]
\begin{center}
\includegraphics[width=1.0\columnwidth]{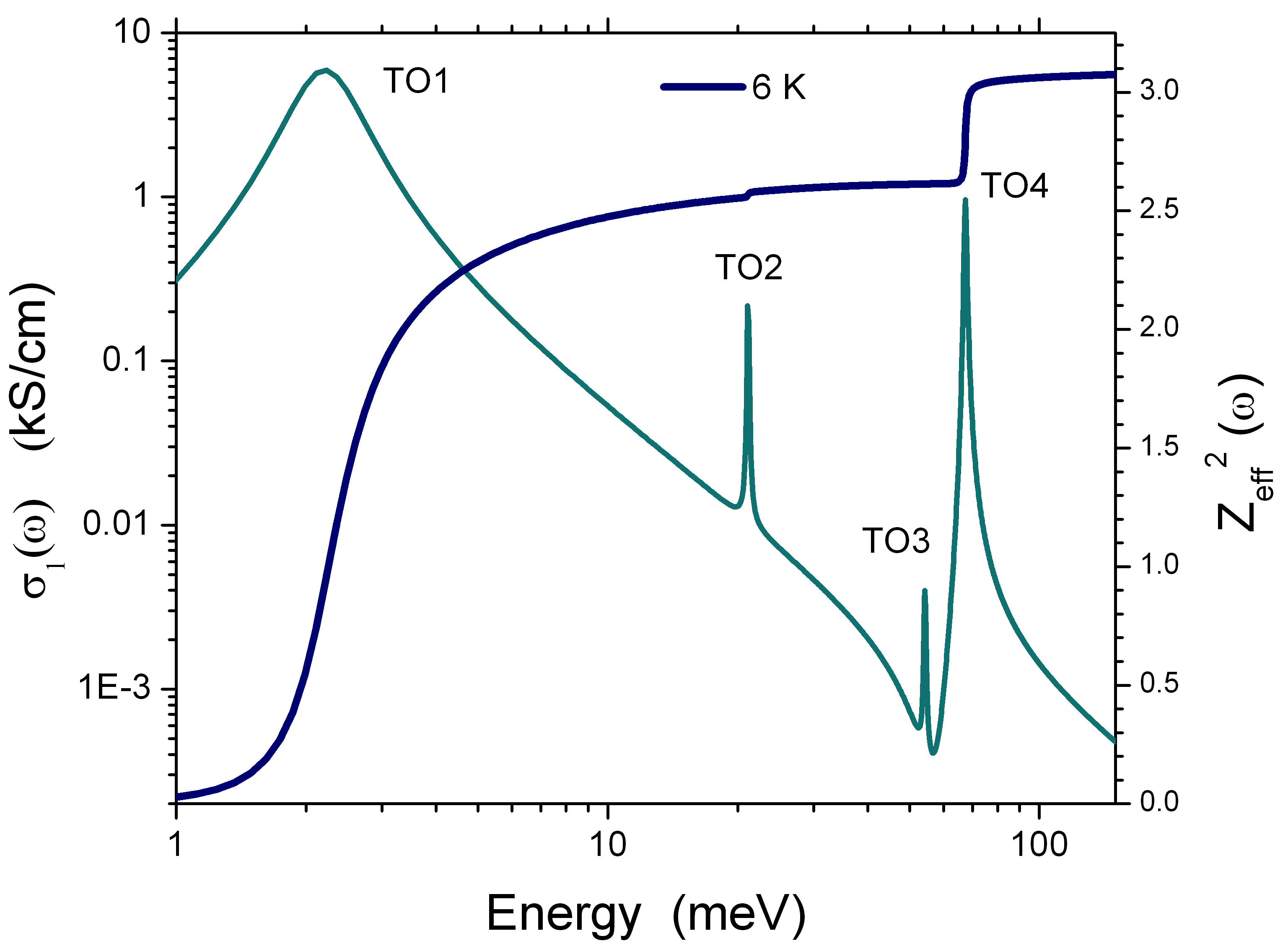}
\caption{\label{fig:sigma-Z-ph}
Optical conductivity $\sigma _1 \left( \omega  \right)$ at 6 K (see Appendix) and the sum-rule integral $Z_{eff}^2$ defined in Eq.~\ref{eq:Zeff}, representing the transverse effective charge of the optical phonons of pristine SrTiO$_3$. }
\end{center}
\end{figure}

In the present manuscript an affirmative answer is provided to this question, but the nature of the interaction is unusual and the pairing process involves the exchange of two transverse polarized phonons as proposed by Ngai~\cite{ngai1974}.
We begin by highlighting a striking feature~\cite{zhong1994,woelfle2018} of the optical phonons of pristine SrTiO$_3$, namely the fact that the TO1 mode has an unusually strong oscillator strength. To make this quantitative we take a look at the integrated spectral weight of all phonons. In a purely ionic sample the integrated phonon spectral weight satisfies the f-sum rule for the different species of ions in the compound, labeled by $j$:
\begin{equation}
\int\limits_0^\infty  {{\sigma _1}\left( {\omega} \right)} d\omega = 
\sum_j\frac{\pi n_j q_j^2}{2 m_j}
\label{eq:fsum}
\end{equation}
The parameters $m_j$, $q_j$ and $n_j$ are the corresponding mass, charge and volume densities and $\sigma _1\left( {\omega} \right)$ is the optical conductivity. The transverse effective charge (or Born charge) is obtained by taking the ratio of both sides of the above expression
\begin{equation}
Z_{eff}^2 = \left[\sum_j\frac{\pi n_j q_j^2}{2 m_j}\right]^{-1} \int\limits_0^\infty  {{\sigma _1}\left( {\omega} \right)} d\omega 
\label{eq:Zeff}
\end{equation}
and is equal to $1$ for a perfectly ionic compound such as MgO~\cite{marel_lecture_notes}. Fig.~\ref{fig:sigma-Z-ph} depicts $\sigma _1\left( {\omega} \right)$ of SrTiO$_3$ at 6 K determined experimentally from reflectivity measurements (see Appendix) and the resulting $Z_{eff}^2$. The latter amounts to a value of $3.1$, implying that the effective charge is a factor $1.8$ enhanced. Since the oxygen ions are by far the lightest ions and therefore are the dominant contributions to Eq.~\ref{eq:fsum}, one would be lead to conclude that the ionic charge is $-3.5$  instead of $-2$, which makes absolutely no sense from a chemical perspective. For the case of  SrTiO$_3$  this was answered in Ref.~\cite{zhong1994}, namely the "Ti-O ionic bond is on the verge of being covalent, leading to large charge transfers"~\cite{woelfle2018}. In 1977, Michael J. Rice has analyzed this phenomenon in a different context of organic conductors, under the banner "charged phonons"~\cite{rice1977}, and the formalism has since then been applied to a number of different cases, including buckminster-fullerene~\cite{rice1992}, FeSi~\cite{damascelli1997} and bi-layer graphene~\cite{cappelutti2012}.

\begin{figure}[t!!]
\begin{center}
\includegraphics[width=1.0\columnwidth]{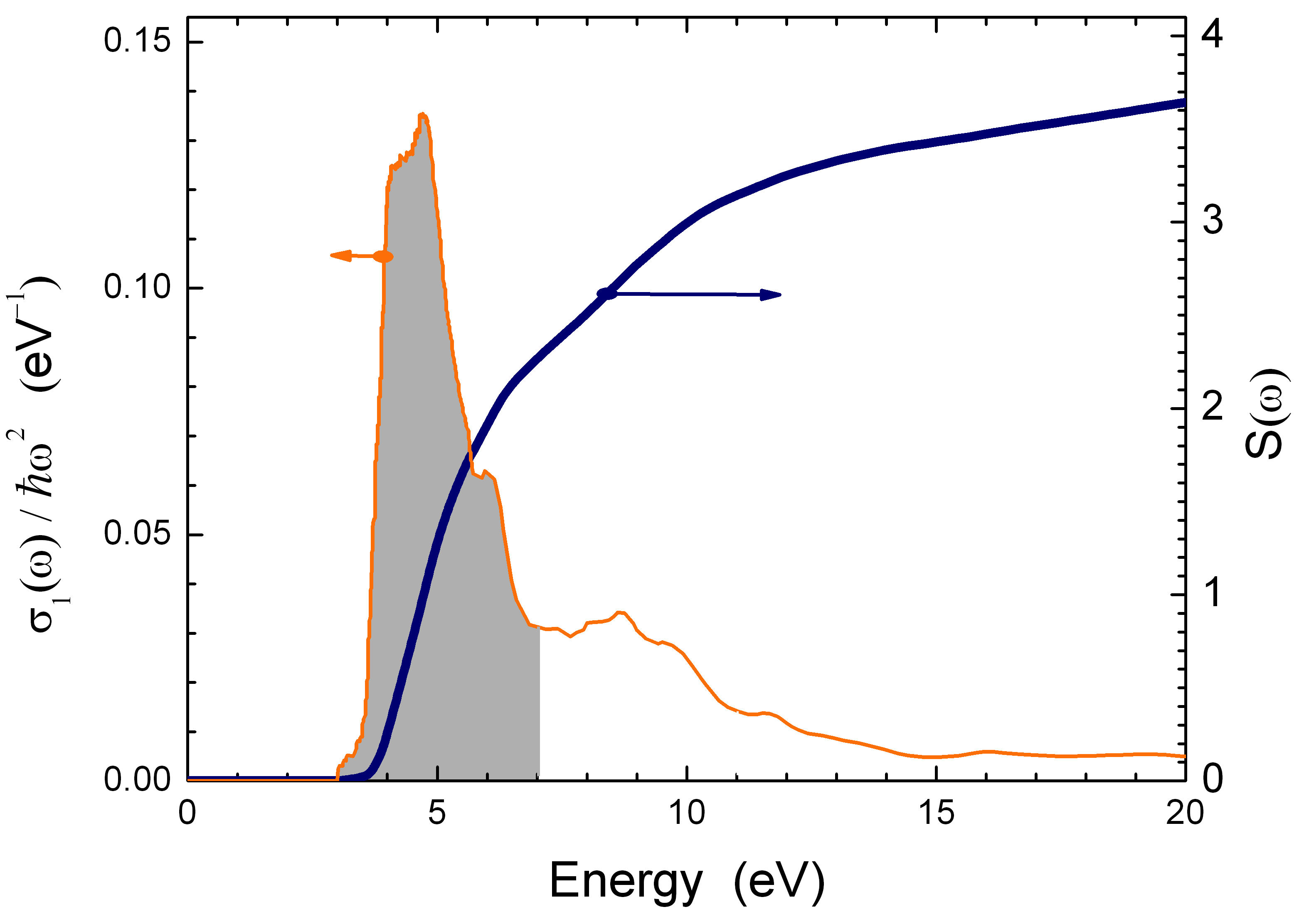}
\caption{\label{fig:sigma-w2-S}
Differential oscillator strength $\sigma(\omega)/\omega^2$  (orange). For $\hbar\omega < 5.7 eV$ the dielectric function of Ref.~\cite{cardona1968} was digitized, the optical conductivity data above 5.7 eV were digitized from Ref.~\cite{asmara2014}. The differential oscillator strength is the contribution of the optical conductivity per unit of energy to the static dielectric permeability $\epsilon(0)$. The dark blue curve represents the integrated oscillator strength taking 3 eV as the lower threshold. The grey area under the orange curve represents the energy range of O$2p$ to Ti$3d(t_{2g})$ transitions. The corresponding value of the oscillator strength of this transition can be read on the righthand axis, corresponding to $S_e=2.2$ eV.  }
\end{center}
\end{figure}

\begin{figure}[t!!]
\begin{center}
\includegraphics[width=0.4\columnwidth]{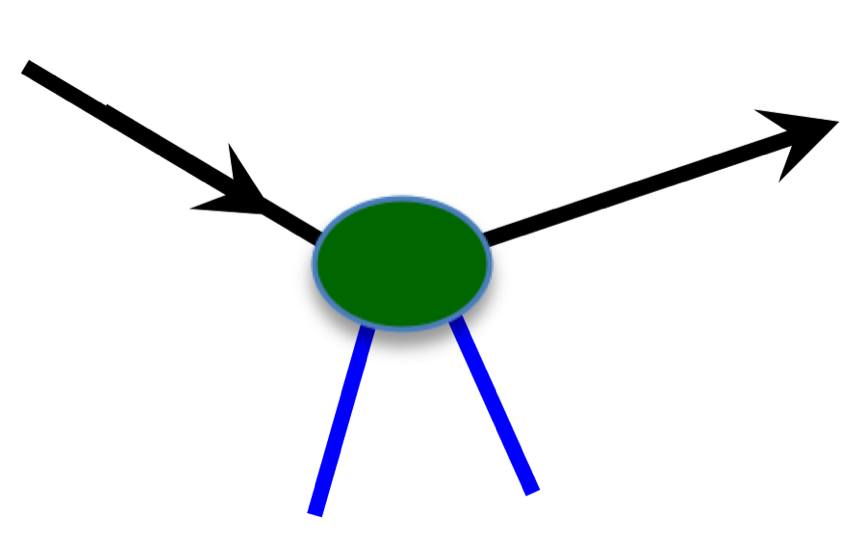}
\includegraphics[width=0.6\columnwidth]{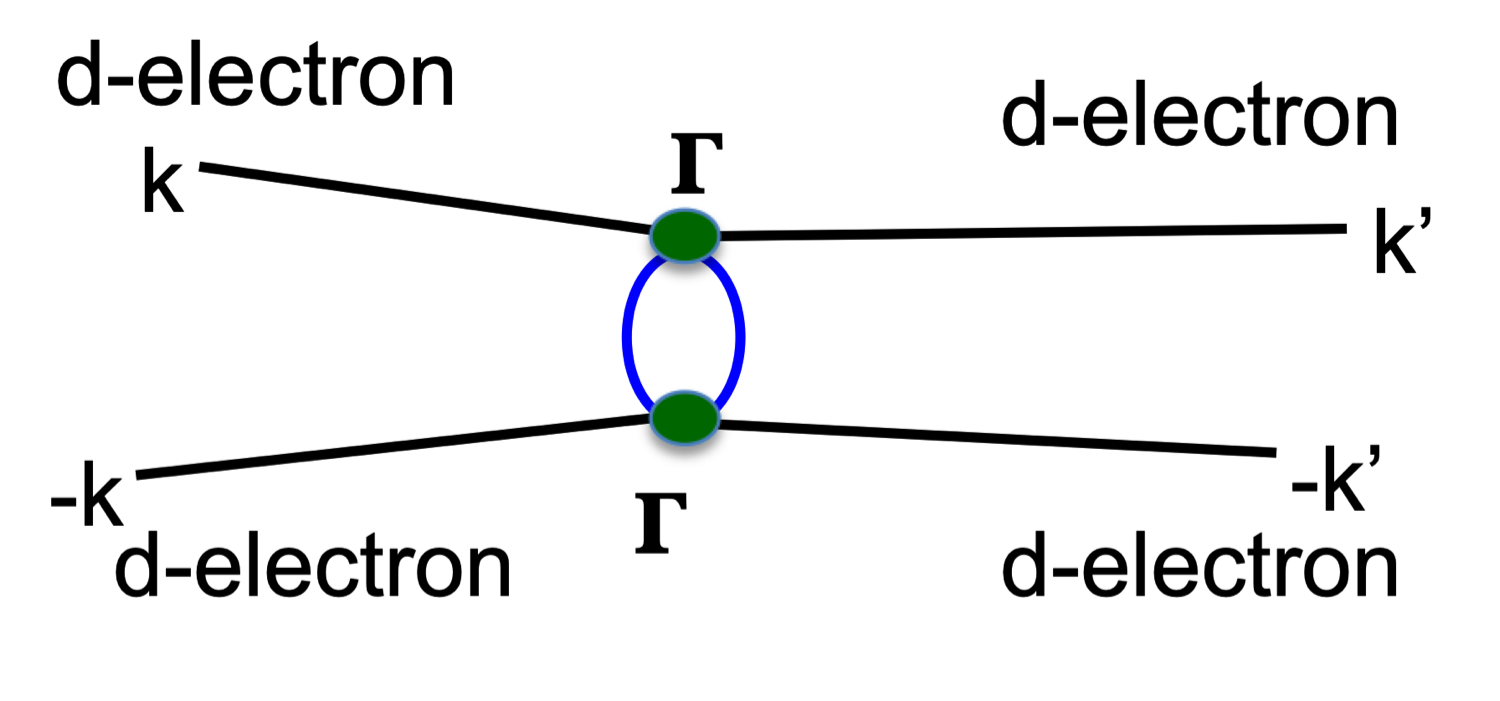}
\caption{\label{fig:diagrams}
Top: Two-phonon process resulting from the charged-phonon hamiltonian, Eq.~\ref{eq:e-ph}, in second order perturbation theory, when the $p$-band is fully occupied. 
Bottom: Two-phonon scattering process involving incoming and outgoing pairs of electrons with zero total momentum.}
\end{center}
\end{figure}

In order to identify the main source of electron-phonon coupling we take advantage of the relation 
\begin{equation}
\varepsilon\left( 0 \right)=1+8\int\limits_0^\infty  \frac{\sigma _1 \left( \omega  \right)}{\omega ^2} d\omega  
\end{equation}
In Fig.~\ref{fig:sigma-w2-S} we show  $\sigma _1 \left( \omega  \right)/ \omega ^2$  as well as the sum rule integral $S\left( \omega  \right)$ of this quantity. From LDA bandstructure calculations one can identify the range from 3 to 7 eV with transitions from the occupied O$2p$ states to the empty Ti$3d(t_{2g})$ bands~\cite{kahn1964}. The latter bands, which are empty in pristine SrTiO$_3$, become populated when electron-doping the material, and the electrons in these bands are those which exhibit superconductivity. The data makes abundantly clear that the lion's share of the static polarizability originates from the mixing of O$2p$ and Ti$3d(t_{2g})$ character, and that the corresponding oscillator strength is $S_e=2.2$.
We can combine this with our knowledge of $Z_{eff}^2$ to obtain the value of the classical (spring) coupling constant $\gamma$, the electronic and vibrational length scales $a_e$  (Eq.~\ref{eq:ae}) and $a_n$ (Eq.~\ref{eq:an}) and the transverse electron-phonon coupling constant $g$ (Eq.~\ref{eq:g}) (see Appendix). The resulting values of these parameters are given in Table~\ref{table1}.
Equipped with these parameters we are now in a position to analyze further consequences of the electron-phonon coupling described by the Hamiltonian (see Appendix):  
\begin{eqnarray}
\hat H&=&\frac{{\hbar {\omega_e}}}{2}\left( {\hat d}^\dag \hat d - {\hat p}^\dag \hat p\right) + \hbar {\omega_n}\left( {\frac{1}{2} + {{\hat a}^\dag }\hat a} \right) 
\nonumber
\\
&-& g\left( {{{\hat d}^\dag }\hat p + {{\hat p}^\dag }\hat d} \right)\left( {{{\hat a}^\dag } + \hat a} \right)
\label{eq:e-ph}
\end{eqnarray}
The momentum dependence is provided in the Appendix. For the present purpose we will just work out the consequences in the local model described by Eq.~\ref{eq:e-ph}.
The first important point to note is the fact that the emission or absorption of a phonon implies a change of parity of the electronic wavefunction. For bands close to the Brillouin-zone center this implies that the electron has to switch necessarily from one band to another. The main players are the occupied O$2p$ bands and the empty Ti$3d$ bands, the latter ones of $t_{2g}$ character immediately above the insulator gap (3 eV) and the $e_g$ states at 2.5 eV higher energy. 
The implication is, that a process whereby a single TO1 phonon is exchanged between two d-electrons has zero amplitude. The next available process involves the exchange of two phonons. We consider first a state with a single $d$ electron doped into the insulator. The electron-phonon interaction in Eq.~\ref{eq:e-ph} generates on top of this a virtual electronic excitation and a phonon, together having an energy $\omega_e+\omega_n$, which decays into a state with again a single $d$-electron and two phonons as shown in Fig.~\ref{fig:diagrams}. In second order perturbation theory the corresponding scattering amplitude is
\color{black}
\begin{equation}
\Gamma \approx - \frac{g^2}{\hbar\omega_e}
\end{equation}
where we used that $\omega_n\ll \omega_e$. The value is reported in Table~\ref{table1}. As depicted in the bottom panel of Fig.~\ref{fig:diagrams} the two emitted phonons can be reabsorbed by a second $d$-electron in a similar process, causing the two $d$-electrons to interact by virtual exchange of two optical phonons. Following the standard treatment of BCS theory we obtain for the product of pairing interaction and density of states $D_F^*$ at $\epsilon_F$:
\begin{equation}
\lambda_{2ph}=2D_F^*\frac{\Gamma^2}{2\hbar\omega_n}
\end{equation}
where the factor 2 in front of the expression accounts for the two possible permutations of the bi-phonon exchange \cite{referee} and the factor two in the denominator accounts for the fact that the interaction is mediated by a pair of phonons.
Combining all experimental factors used to evaluate $g$,  $\Gamma$ and $\lambda_{2ph}$ we can write this as
\begin{equation}
\lambda_{2ph}=\frac{
\hbar
(8\pi)^2
n_{e,p}^2
q_e^4
\left(Z_{eff}^2-1\right)^4
D_F^*
}{
S_e^2
m_n^2
\omega_n^3
}
\end{equation}
The thus obtained value for the electron-electron coupling constant is $\lambda_{2ph}=0.28$.
Combining the BCS equation for the critical temperature 
\begin{equation}
k_BT_c=2\hbar\omega_ne^{-1/\lambda}
\end{equation}
(where the only difference is, that we assume here that pairing is mediated by bi-phonons, thus doubling the prefactor) with the experimental value of $T_c\le 400$ mK, we arrive at  $\lambda_{exp}=0.15$. The small value $\lambda$ carries over to a huge amplification of uncertainty in the value of $T_c$. It is therefor not useful to compare theoretical and experimental $T_c$'s. However, the close value of theoretical and experimental $\lambda$ is encouraging and shows that Ngai's model~\cite{ngai1974} of the exchange of two optical phonons could be relevant for superconductivity in SrTiO$_3$. 
\begin{center}
    \begin{table}
        {\small
        \hfill{}
        \begin{tabular}{|c|c|c|l}
        \hline
        quantity & value & units \\
        \hline
        $Z_{eff}^2$& 3.1&  ~$[a]$\\ 
        $N_{e,p}$ & 2.0 &  ~$[a]$\\        
        $V$ & $5.95\cdot 10^{-23}$ & cm$^{3}$ ~\cite{koonce1967}  \\ 
        $\hbar\omega_n$& 12 & meV~$[b]$\\   
        $\hbar\omega_e$& 5.0 & eV~$[a]$\\
        $S_e$& 2.2& ~$[a]$\\        
        $D^*_F$& $\le 0.80$&  eV$^{-1}$~\cite{marel2011}\\   
        $T_c$& $\le 400$ &  mK~\cite{schooley1964}\\          
        \hline
        \end{tabular}}  
        {\small   
        \begin{tabular}{|c|c|c|l}   
        \hline        
        quantity & value & units \\                
        \hline        
        $\tilde{m}_e$& 0.84&  $m_e$\\   
         $a_e$& 9.5$\cdot 10^{-9}$& cm \\   
        $a_n$& 1.0$\cdot 10^{-9}$&  cm\\   
        $\gamma$&$9.3\cdot 10^{4}$ & g/s$^2$ \\    
        $g$& 0.57& eV \\
        $\Gamma$& 66&  meV\\       
        $\lambda_{2ph}$& $\le 0.28$& \\   
        $\lambda_{exp}$& $\le 0.15$& \\                      
        \hline
        \end{tabular}}
        \hfill{}
        \caption{List of the experimental quantities (left) and derived theoretical quantities (right).
        $[a]$ Present work.   $[b]$ Estimated from the $\omega(q)$ dispersion of TO1 reported in Ref.~\cite{stirling1972}    }
        \label{table1}
    \end{table}
\end{center}

Ngai pointed out another important implication of the fact that this pairing process involves two phonons~\cite{ngai1974}: the main reason why acoustic phonons are ineffective as pairing mediators in ultra low density superconductors such as SrTiO$_3$, is the fact that the BCS pairing interaction scatters a $k,-k$ electron pair into a $k+q$,$-k-q$ pair with $q$ the phonon momentum: The Fermi-momentum provides a constraint on $q$ in momentum space, thereby limiting the phonon energies to values smaller than $2c k_F$ where $c$ is the sound velocity. This phase space constraint restricts the available phonon energies to a small fraction of the Debye energy, thereby limiting $T_c$. These phase space constraints do not apply to 2-phonon processes, since two phonons of approximately opposite momentum can individually have an arbitrary high momentum and energy.

{\color{black} We started out by analyzing the observed, strong, charged phonon effect in pristine SrTiO$_3$, and proceeded to extract an electron-phonon coupling constant relevant in bi-phonon pairing processes. Usually when the charged phonon effect is observed, the electronic oscillator and the phonon oscillator are much closer in energy, and in fact often overlap, leading to interesting resonance behaviour and Fano-asymmetries. In the present case the vibrational and electronic energy scales are separated by two orders of magnitude, making the case of SrTiO$_3$ quite exceptional in that the large observed $Z_{eff}$ implies that the charged phonon interaction is abnormally strong. These materials are close to a ferroelectric instability, such an instability requires the harmonic potential to switch sign near the phase transition, the ferroelectric displacement is exactly what is driving the charged phonon effect, and it is difficult to identify a pairing mechanism of the doped superconducting materials. From an intellectual point of view it would be satisfying if a common factor could be identified which is responsible for all of those phenomena. The model proposed in the present manuscript fits this description, and solves the conundrum of the electron coupling to the soft ferroelectric modes despite the fact that these modes are transverse polarized. The softening of the mode brings along the difficulty that any anharmonicity in the potential of the vibrational coordinate becomes particularly pronounced ({\em e.g.} a mexican hat shape) and the vibrational frequency becomes imaginary~\cite{aschauer2014}. This immediately leads to the problem that parameters such as $\gamma$ and $a_n$ lose their meaning at least in the long wavelength limit where the softening takes place. A treatment of the bi-phonon exchange processes including aforementioned features should be subject of future studies. The current state of affairs provides a motivation to undertake further theoretical and experimental studies. At the present stage a number of falsifiable consequences can already be stated. In particular the soft two-phonon processes, corresponding to a broad continuum of the bi-phonon energies (about 25 to 30 meV), should show up in tunnelling and photoemission spectra, with an electron-phonon coupling constant $g\sim 0.6$ eV. Another consequence is, that pulsed optical excitation of the $pd$ electron-hole excitations using a laser operating between 3 and 7 eV, should lead to pronounced coherent excitation of the soft ferroelectric modes. This method provides an alternative scheme to measure the coupling constant $g$.}  

This project was supported by the Swiss National Science Foundation (project 200021-162628). DvdM acknowledges insightful discussions with Dung-Hai Lee, Dimitri Abanin, Nicola Spaldin, Alexander Balatsky, Jonathan Ruhman, Thom Devereaux, and Jan Zaanen. 

\section{Appendix}
\subsection{Details of the formalism}
\label{subsection:formalism}

Bringing elements together from scattered texts comes with the risk of missing a factor of $\pi$ or $1/2$ here and there. Since at the end of the day we need to compute numbers that can be compared with experiments we will derive the charged-phonon formalism and the resulting bi-phonon pairing mechanism from scratch. 

\begin{figure}[h!!]
\begin{center}
\includegraphics[width=1\columnwidth]{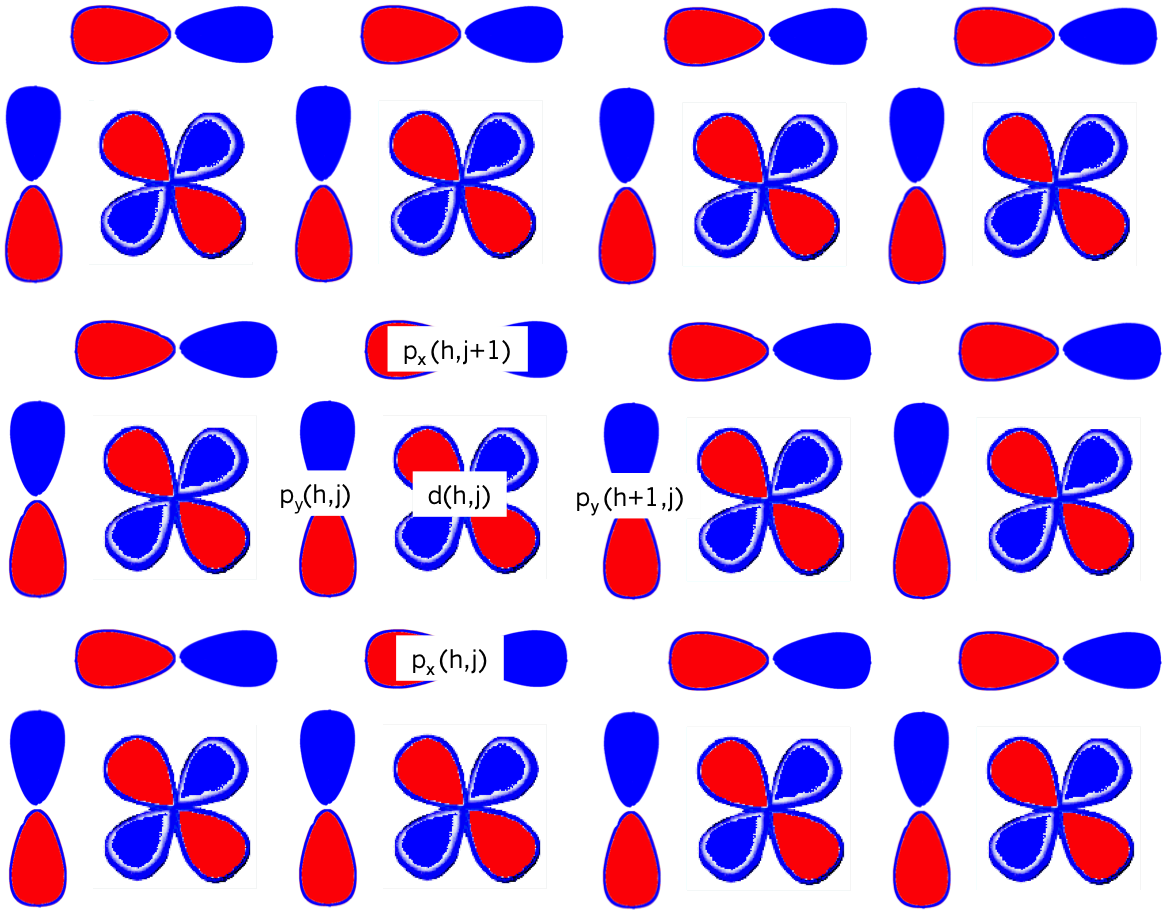}
\caption{\label{fig:orbitals}
While SrTiO$_3$ is a three dimensional crystal, the relevant valence states are built up from oxygen 2p and titanium 3d$_{xy}$ orbitals as sketched in this figure. If we label horizontal and vertical axis as $x$ and $y$, one should imagine that this structure is repeated along the z-axis of the simple cubic structure. The corresponding bands are degenerate with, and orthogonal to, the bands comprised of 3d$_{yz}$ and 3d$_{zx}$ with corresponding oxygen $2p$ orbitals. }
\end{center}
\end{figure}

We consider a two-level system of volume $V$ with occupied $p$-levels and empty $d$-levels, which we assume to be centered at the same atomic site. The relevant electronic levels are depicted in Fig.~\ref{fig:orbitals}. For two O$^{2-}$ ions adjacent to Ti$^{4+}$ ions with the three ions aligned along the $x$ axis this implies that we consider the "bonding" molecular orbital of the two O $p_y$ orbitals with its center of mass at the Ti $3d_{xy}$ state.
For electric field polarized along the x-direction, the p-d transition is dipole allowed.
This system is described by the Hamiltonian:
\begin{equation}
\hat{H}_e = \frac{{\hbar {\omega_e}}}{2}\left( {\hat d}^\dag \hat d - {\hat p}^\dag \hat p\right)
\end{equation}
The position operator follows from the consideration that $p$ and $d$ have the same center of mass. A linear combination $u \left|p\right\rangle + v \left|d\right\rangle$ will have its center of mass displaced in proportion to $uv$. The position operator should therefore be of the form
\begin{equation}
{{\hat x}_e} = {a_e}\left( {{{\hat d}^\dag }\hat p + {{\hat p}^\dag }\hat d} \right)
\end{equation}
where $a_e$ is a length scale that we will obtain later from considerations of the optical sum-rule for the $p-d$ transition. The velocity operator then follows from the Heisenberg equation of motion:
\begin{equation}
{{\hat {\dot {x}}}_e} = \frac{i}{\hbar}\left[ {\hat{H}_e,{{\hat x}_e}} \right] =
i{a_e}{\omega _e}\left( {{{\hat d}^\dag }\hat p - {{\hat p}^\dag }\hat d} \right)
\end{equation}
Applying the Heisenberg equation of motion a second time provides the result that the position operator satisfies the differential equation of a harmonic oscillator:
\begin{equation}
{{\hat {\ddot {x}}}_e} = \frac{i}{\hbar}\left[ \hat H_e,\hat{\dot {x}}_e \right]
=- \omega_e^2{{\hat x}_e}
\end{equation}
For later use it is useful to work out the commutator of the position and the velocity operator. By means of straightforward operator algebra we obtain
\begin{equation}
\left[ {{{\hat x}_e},{{\hat {\dot {x}}}_e}} \right] = 
-\frac{{4ia_e^2}}{\hbar}\hat{H}_e
\end{equation}
To continue beyond this point we need to specify the groundstate to which we are going apply these rules. Here we consider the case that the $d$-state is empty, and the $p$-state is occupied, {\em i.e.}
$\left\langle {\hat d}^\dag \hat d \right\rangle  = 0$ \& $\left\langle {\hat p}^\dag \hat p \right\rangle=N_{p,e} $, than:
\begin{equation}
\left\langle \hat H_e \right\rangle  =  - N_{p,e}\frac{\hbar \omega_e}{2}
\end{equation}
so that
\begin{equation}
\left\langle\left[ {{{\hat x}_e},{{\hat {\dot {x}}}_e}} \right]\right\rangle 
=2ia_e^2{\omega_e}N_{p,e}
\label{eq:commutator}
\end{equation}
{\color{black} From the theory of the linear response of the current to the electric field $E$ it follows that the optical conductivity $\sigma=n_e q_e\dot{x}_e/E$ is, apart from a factor $\omega$, given by the sum of the diamagnetic contribution and the current-current correlation function:
\begin{equation}
\sigma \left( \omega  \right) = 
\frac{q_e^2 \left\langle \left[ \hat{x}_e,\hat{\dot{x}}_e \right]\right\rangle }{\hbar \omega V}
 - \frac{q_e^2}{\hbar \omega V}\int\limits_0^\infty  
     e^{i\omega t}
    \left\langle \left[ \hat{\dot{x}}_e,\hat{\dot{x}}_e(t) \right]\right\rangle
   dt  
\label{sigma}
\end{equation}
Integration over all frequencies gives the generic commutator expression:
\begin{equation}
\int\limits_0^\infty \sigma _1 \left( \omega  \right)d\omega  
=
\frac{{ \pi {q_e^2}}}{{2\hbar V}} \mbox{Im}
 \left\langle \left[\hat{{x}}_e, \hat{\dot{x}}_e\right]\right\rangle 
\end{equation}
After substitution of the result of the commutator obtained above (Eq.~\ref{eq:commutator}) we arrive at
\begin{equation}
\int\limits_0^\infty  \sigma _1\left( \omega  \right)d\omega  
=  \frac{\pi n_{p,e} q_e^2 a_e^2 \omega_e}{\hbar} = \frac{\pi n_{p,e} q_e^2}{2\tilde{m}_e} 
\label{eq:f-sum}
\end{equation}
where $n_{p,e}$ is the volume density of the $p$-electrons ({\em i.e.} the number of electrons in the $2p$-shell coupled to the Ti-atom divided by the cell volume $V$). The rightmost expression resembling the f-sum rule constitutes in fact the {\em definition} of the effective mass $\tilde{m}_e$ of the electrons excited in the $p-d$ transition. While in the present case the numbers work out to provide a mass value close to that of a free electron, the value of $\tilde{m}_e$ employed in the analysis of the present model has to be taken from Eq.~\ref{eq:f-sum} in the interest of internal consistency of the formalism. }

Integration of $ \sigma _1( \omega)/\omega^2 $ provides another useful quantity, the static dielectric permittivity:
\begin{equation}
\int\limits_0^\infty  \frac{\sigma _1 \left( \omega  \right)}{\omega ^2} d\omega  =
\frac{\varepsilon\left( 0 \right) - 1}{8}  =  \frac{S_e}{8} =\frac{\pi n_{p,e} q_e^2 a_e^2}{\hbar \omega_e} 
\end{equation}
The leftmost equality is a consequence of the Kramers-Kronig relation between the dissipative and dispersive components of the dielectric function. The last expression on the right is specific to the present model, and is a consequence of $\sigma_1(\omega)$ being proportional to $\delta(\omega-\omega_e)$. 

By comparing the different members of the two expressions provided above, we see that
the characteristic length scale $a_e$ can be expressed either as a function of $\tilde{m}_e$ or as a function of $S$: 
\begin{equation}
{a_e} = \sqrt {\frac{\hbar}{2\tilde{m}_e\omega_e}} 
= \sqrt {\frac{\hbar {\omega_e}S_e}{\pi n_{p,e}q_e^2}} 
\label{eq:ae}
\end{equation}
Comparing the two righthand members of the above equation we see, that the effective mass for the $p-d$ transition is provided by the expression:
\begin{equation}
\tilde{m}_e \equiv \frac{{4\pi n_{p,e}{q_e^2}}}{{S_e\omega_e^2}}
\end{equation}
Since we wish to extract parameters from optical data using a Lorentz-oscillator model both for the vibrational and electronic component, we use for this part of the analysis a boson representation of the electronic degree of freedom:
\begin{eqnarray}
&\hat{b}^\dag = \hat{d}^\dag  \hat{p}
&
\hat{b} =  \hat{p}^\dag\hat{d}
\nonumber
\\
&{{\hat x}_e} ={a_e}\left( {\hat{b} + {{\hat{b}}^\dag }} \right)
\hspace{1cm}
&
{{\hat{p}}_e} = \frac{\hbar}{{2i{a_e}}}\left( {\hat{b} - {{\hat{b}}^\dag }} \right)
\end{eqnarray}
It is important to note, that the $\hat{b}$ operators don't satisfy Bose-commutation rules. They do however provide a correct representation of the system provided that we restrict the states of the system to zero or one boson. In this limited sense we can map the electronic subsystem on a harmonic oscillator. The normal modes of the coupled system can be obtained by diagonalisation of the coupled electronic-vibrational Hamiltonian in the classical limit. Using the above definitions of the electronic coordinates together with the vibrational coordinates, we obtain the hamiltonian
\begin{eqnarray}
H &=&\frac{{p}_e^2}{2\tilde{m}_e}+\frac{\tilde{m}_e\omega_e^2}{2}{x}_e^2
+\frac{{p}_n^2}{2m_n}+\frac{m_n\omega_n^2}{2}{x}_n^2
\nonumber
\\
&+& E (q_e{{ x}_e}+q_n{{ x}_n}) - \gamma {{ x}_e}{{ x}_n} 
\end{eqnarray}
where $\gamma$ is the spring constant connecting the electronic and vibrational degrees of freedom. In the interest of compactness we have absorbed all terms proportional to $x_e^2$ and $x_n^2$ in the definitions of $\omega_e$ and $\omega_n$.
For a harmonic time-varying electric field with angular frequency $\omega$, the nuclear and electronic currents are
\begin{equation}
{j_n}= i\omega {x_n}q_n{n_n} 
\hspace{3mm}\mbox{}\hspace{3mm}
{j_e}= i\omega {x_e}q_e{n_e} 
\end{equation}
and the coupled classical equations of motion become:
\begin{equation}
\left[ {\begin{array}{*{20}{c}}
{\tilde{m}_e\left( {{\omega ^2} - \omega _e^2} \right)}&\gamma \\
\gamma &{{m_n}\left( {{\omega ^2} - \omega _n^2} \right)}
\end{array}} \right]
\left( {\begin{array}{*{20}{c}}
{{x_e}}\\
{{x_n}}
\end{array}} \right) = \left( {\begin{array}{*{20}{c}}
{q_eE}\\
{q_nE}
\end{array}} \right)
\end{equation}
The solution for the vibrational conductivity $\sigma_n=j_n/E$ is
\begin{equation}
{\sigma _n}\left( \omega  \right) = 
\frac{{{q_n}^2 n_n}}{{m_n }}
\frac{{i\omega Z_{eff}^2\left( \omega  \right)}}{{{\omega ^2} - \tilde \omega_n^2\left( \omega  \right)}}
\end{equation}
Due to the electron-phonon coupling the vibrational frequency has shifted to
 \begin{equation}
\tilde \omega_n^2\left( \omega  \right) = \omega_n^2 - \frac{{{\gamma ^2}m_n^{ - 1}\tilde{m}_e^{ - 1}}}{{\omega_e^2 - {\omega ^2}}}
\end{equation}
The transverse effective charge is given by
\begin{equation}
Z_{eff}^2\left( \omega  \right) = 1 + \frac{{\gamma \tilde{m}_e^{ - 1}}}{{\omega_e^2 - {\omega ^2}}}
\end{equation}
In the case of SrTiO$_3$ the electronic and vibrational frequencies are separated by two orders of magnitude. Consequently, for the range of vibrational frequencies this expression can be used in the limit $\omega\rightarrow 0$. The coupling constant $\gamma$ is readily obtained from the transverse effective charge through the relation
\begin{equation}
\gamma=\left(Z_{eff}^2-1\right)\tilde{m}_e\omega_e^2=    \left(Z_{eff}^2-1\right)\frac{{4\pi n_{p,e}{q_e^2}}}{{S_e}}
\label{eq:gamma}
\end{equation}
Once we have read out $Z_{eff}$, $S_e$, $\omega_e$, $\omega_n$, and $\gamma$ from the optical data, we are ready to set up the observables on the quantum level.
The nuclear position and momentum operators are: 
\begin{equation}
{{\hat x}_n} = {a_n}\left( {\hat a + {{\hat a}^\dag }} \right)
\hspace{3mm}\mbox{}\hspace{3mm}
{{\hat p}_n} =\frac{\hbar}{{2i{a_n}}}\left( {\hat a - {{\hat a}^\dag }} \right)
\end{equation}
where the characteristic length scale is given by
\begin{equation}
{a_n} = \sqrt {\frac{\hbar}{{2{m_{n}}{\omega_n}}}} 
\label{eq:an}
\end{equation}
The Hamiltonian operator of the coupled electron-phonon system is:
\begin{eqnarray}
\hat H&=&\frac{{\hbar {\omega_e}}}{2}\left( {\hat d}^\dag \hat d - {\hat p}^\dag \hat p\right) + \hbar {\omega_n}\left( {\frac{1}{2} + {{\hat a}^\dag }\hat a} \right) 
\nonumber
\\
&-& g\left( {{{\hat d}^\dag }\hat p + {{\hat p}^\dag }\hat d} \right)\left( {{{\hat a}^\dag } + \hat a} \right)
\end{eqnarray}
The constant $g$ is the spring constant connecting the electron and nuclear oscillator, multiplied by the characteristic length of the electronic oscillator and the nuclear one:
\begin{equation}
g \equiv \gamma {a_n}{a_e}
\label{eq:g}
\end{equation}
\subsection{Momentum dependence of bandstructure and electron-phonon interaction}
\label{subsection:bandstructure}
The relevant orbitals are depicted in Fig.~\ref{fig:orbitals}
The tightbinding hamiltonian with nearest neighbor hopping corresponding to the orbitals in a single plane is:
\begin{eqnarray}
\hat{H} &=& \sum_{h,j}
\left(
\epsilon_d \hat{d}^{\dagger}_{xy,h,j}\hat{d}_{xy,h,j}+\epsilon_p \hat{p}^{\dagger}_{x,h,j}\hat{p}_{x,h,j}+\epsilon_p \hat{p}^{\dagger}_{y,h,j}\hat{p}_{y,h,j}
\right) 
\nonumber\\
&+&
t_{pd}\sum_{h,j} 
\left(\hat{p}^{\dagger}_{y,h,j}\hat{d}_{xy,h,j}+\hat{p}^{\dagger}_{x,h,j}\hat{d}_{xy,h,j}\right)  + {\mbox h.c.} 
\nonumber\\
&-&
t_{pd}\sum_{h,j} 
\left(\hat{d}^{\dagger}_{xy,h,j}\hat{p}_{y,h+1,j}+\hat{d}^{\dagger}_{xy,h,j}\hat{p}_{x,h,j+1}\right)  - {\mbox h.c.} 
\label{eq:tbham}
\end{eqnarray}
Diagonalization is straightforward and gives
\begin{equation}
\hat{H}= \sum_{j=1}^3\sum_{k} 
\epsilon_{k,j} \hat{c}^{\dagger}_{k,j}\hat{c}_{k,j}
\label{eq:tbham_diag1}
\end{equation}
where the three bands are described by the following energy-momentum dispersions:
\begin{eqnarray}
\epsilon_{k,1}&=&\frac{\epsilon_d+\epsilon_p}{2}-
\sqrt{\frac{(\epsilon_d-\epsilon_p)^2}{4}+\tau_{\vec{k}}^2 }
\nonumber\\
\epsilon_{k,2}&=&\epsilon_p
\nonumber\\
\epsilon_{k,3}&=&\frac{\epsilon_d+\epsilon_p}{2}+
\sqrt{\frac{(\epsilon_d-\epsilon_p)^2}{4}+\tau_{\vec{k}}^2}
\nonumber\\
\tau_{\vec{k}}&=&2t_{pd}\sqrt{\sin^2\left(\frac{k_x a}{2}\right)+\sin^2\left( \frac{k_y a}{2}\right)}
\label{eq:tbham_diag2}
\end{eqnarray}
The natural cause of the electron-phonon interaction in this manuscript is the Peierls interaction, the  modulation of the bond-strength due to the lattice displacement, which leads to the following expression:
\begin{eqnarray}
 \hat{H}_c =  - g'\sum\limits_{h,j}  \hat{d}_{h,j}^\dag \left( \hat{p}_{y,h,j}+ \hat{p}_{y,h + 1,j} \right)
\left(\hat{a}_{x,h,j}^\dag  + \hat{a}_{x,h,j} \right) \nonumber \\
 - g'\sum\limits_{h,j} 
 \hat{d}_{h,j}^\dag \left( \hat{p}_{x,h,j} + \hat{p}_{x,h,j + 1} \right)
\left( \hat{a}_{y,h,j}^\dag  + \hat{a}_{y,h,j} \right)   - {\mbox h.c.} 
\end{eqnarray}
Fourier transformation gives for the special case of phonons travelling along $q_y$ ( $q_x=0$):
\begin{eqnarray}
\hat{H}_c &=&  - 2g'\sum\limits_{\vec{k},q_y}
\cos \left( \frac{k_x a}{2} \right)
\left[
       \hat{d}_{\vec{k}}^{\dag}  \hat{p}_{y,\vec{k} + \vec{q}} 
   +  \hat{p}_{y,\vec{k} - \vec{q}}^{\dag}  \hat{d}_{\vec{k}}
\right]
\left[ \hat{a}_{x,\vec{q}}^{\dag}  +  \hat{a}_{x,-\vec{q}} \right]
\nonumber \\
 &-& 2g'\sum\limits_{\vec{k},q_y}
  \cos \left( \frac{k_y a+ q_y a}{2} \right) 
  \hat{d}_{\vec{k}}^{\dag}  \hat{p}_{x,\vec{k} + \vec{q}} 
\left[ \hat{a}_{y,\vec{q}}^{\dag}  +  \hat{a}_{y,-\vec{q}}  \right]  
\nonumber \\
 &-& 2g'\sum\limits_{\vec{k},q_y}  
  \cos \left( \frac{k_y a- q_y a}{2} \right) 
  \hat{p}_{x,\vec{k} - \vec{q}}^{\dag}  \hat{d}_{\vec{k}}
\left[ \hat{a}_{y,\vec{q}}^{\dag}  +  \hat{a}_{y,-\vec{q}}  \right]
\end{eqnarray}
The operators $\hat{d}_{\vec{k}}^{\dag}$, $\hat{p}_{x,\vec{k}} $ and $\hat{p}_{x,\vec{k} }$ can be decomposed in the band-eigenstate operators $\hat{c}_{\vec{k},j}^{\dag}$. For the present purpose it suffices to point out that the two main players are bands $j=2$ and $j=3$. The character of band $2$ is a pure non-bonding p-band. In the limit of large $\omega_e/t_{pd}$ the character of band $j=3$ becomes purely Ti$3d$. For the purpose of the present paper, where the coupling constants are obtained from experimental data, it is not necessary to work out these equations in more detail. It is however interesting and of potential importance to notice that the Peierls coupling provides in addition to the charged phonon term (the first one) also a coupling to longitudinal phonons (second and third lines of the equation).

\section{Experimental optical conductivity data, energy loss function and electron-phonon coupling function}
\label{subsection:frohlich}
\begin{figure}[b!!]
\begin{center}
\includegraphics[width=1.0\columnwidth]{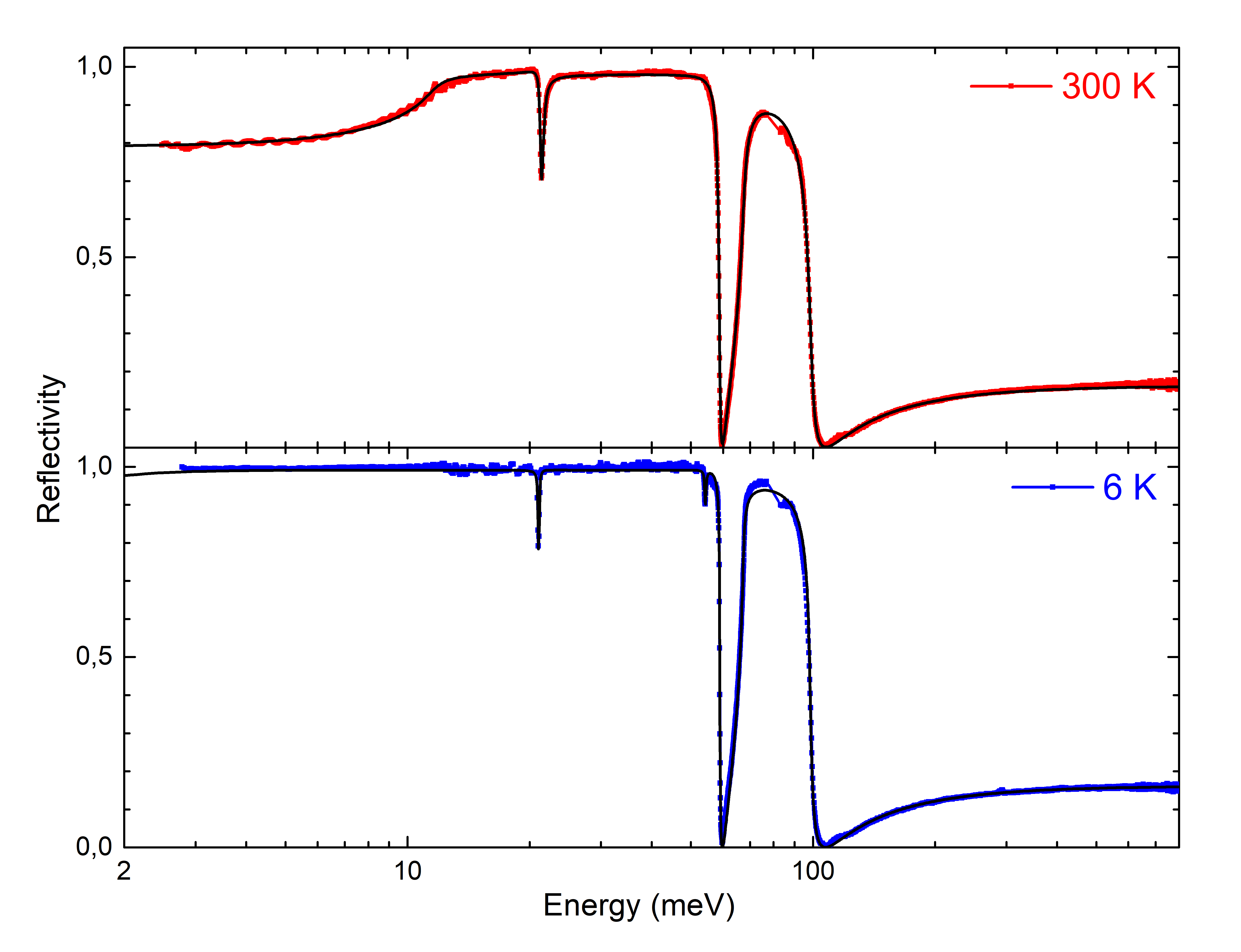}
\caption{\label{fig:reflectivity-fit}
Experimental reflectivity data of pristine SrTiO$_3$ at room temperature and 6 K and fits (solid black curves) using Drude-Lorentz oscillators with Fano-shape asymmetry.}
\end{center}
\end{figure}
\begin{figure}[t!!]
\begin{center}
\includegraphics[width=1.0\columnwidth]{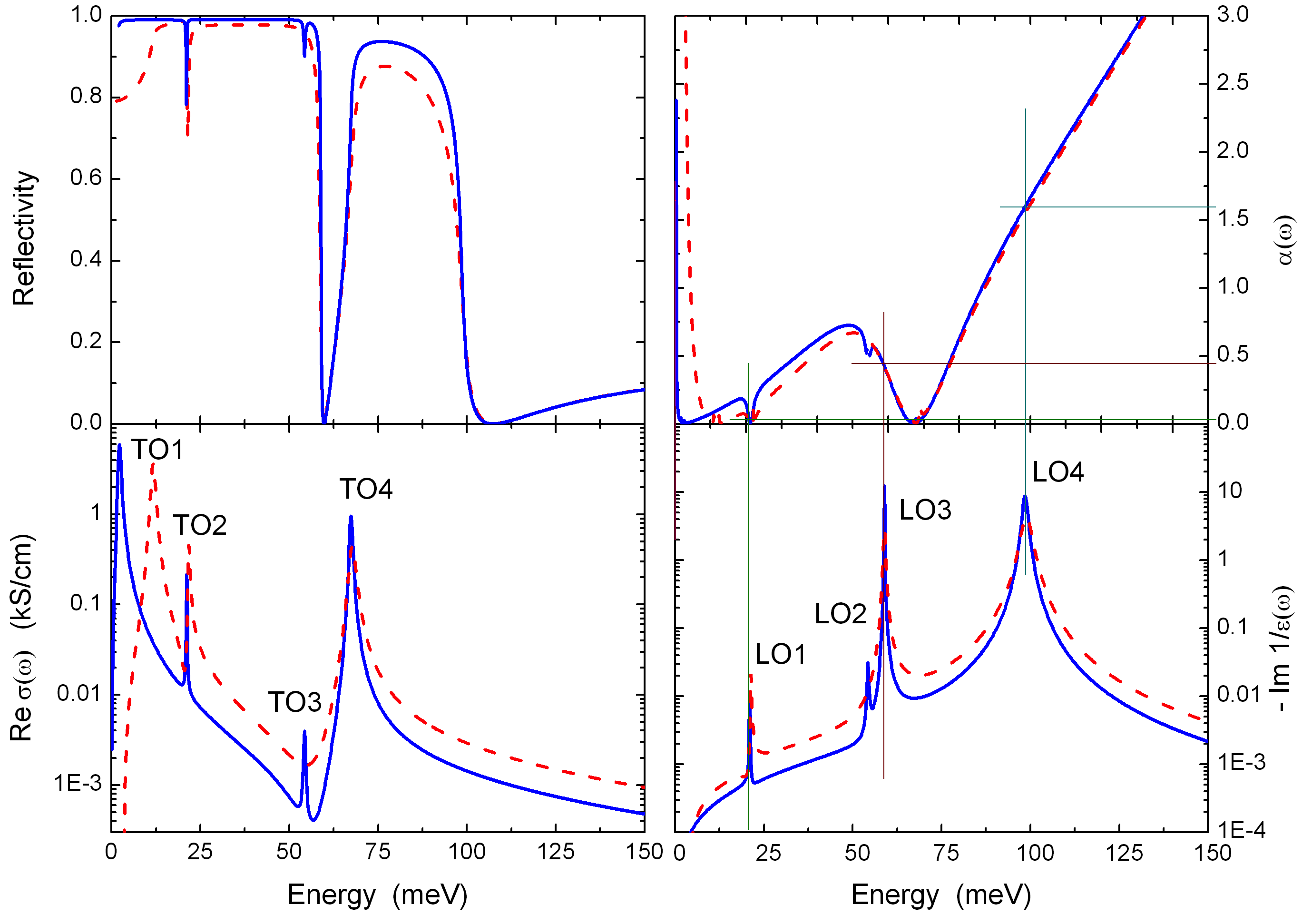}
\caption{\label{fig:longitudinal-coupling}
Reflectivity (top left), optical conductivity (bottom left), loss function (bottom right) and coupling function (top right) of pristine SrTiO$_3$ at room temperature (red dashed) and 6 K (solid blue) curve. The peaks of the optical conductivity (loss function) correspond to the TO (LO) modes. To allow for $\alpha(\omega)$ to be smooth despite the frequency derivative in the denominator, a Lorentz parametrization of the optical spectrum was used. The vertical and horizontal lines in the right hand panels graphically represent the $\alpha$ coefficients corresponding to each of the longitudinal phonons.}
\end{center}
\end{figure}
Fig.~\ref{fig:reflectivity-fit} shows the experimental reflectivity spectra of SrTiO$_3$ at room temperature and 6 K. The data has been fitted with a multi-oscillator function. The complex dielectric function obtained by this procedure is a smooth function suitable for calculating the derivative in Eq.~\ref{toyozawa}. The corresponding optical conductivity shown together with the loss function in Fig.~\ref{fig:longitudinal-coupling} confirms the results reported in Refs.~\cite{servoin1980,kamaras1995,mechelen2008}. The peaks in the optical conductivity coincide with the TO modes, the peaks in the loss function with the LO modes. The top right panel shows the electron-phonon coupling constants calculated using Eq.~\ref{toyozawa}. 
The first point to notice is the strong difference in temperature dependence of TO and LO modes. Whereas the TO1 mode is strongly temperature dependent and approaches zero energy at zero temperature, all LO modes including the one lowest in energy (LO1) are essentially temperature independent. 

The second observation is that the electron-phonon coupling to the LO1 mode is negligible as compared to the other two. This is seen from the coupling constants described by Toyozawa's multi-phonon generalization of the Frohlich model~\cite{frohlich1954,toyozawa1972},
\begin{equation}
{\alpha _j} = \sqrt {\frac{{2{m_b}}}{{{\hbar ^3}\omega _{L,j}^3}}} \frac{{{q_e^2}}}{{{{\left[ {{{\partial \varepsilon } \mathord{\left/
 {\vphantom {{\partial \varepsilon } {\partial \omega }}} \right.
 \kern-\nulldelimiterspace} {\partial \omega }}} \right]}_{L,j}}}}
 \label{toyozawa}
\end{equation}
The values of $\alpha$ can be readily read off from the top right panel of Fig.~\ref{fig:longitudinal-coupling}, and correspond to those that have been used by Devreese {\em et al.}~\cite{devreese2010} to calculate the mid-infrared band of electron-doped SrTiO$_3$. The good agreement of those calculations with the data of Ref.~\cite{mechelen2008} demonstrates the validity of the $\alpha$ parameters obtained with the method of Frohlich. We should therefore take seriously that the electron-phonon coupling to the lowest energy longitudinal mode is very small, and most likely insignificant in the context of the mechanism for superconductivity.
%
%

%
\end{document}